\documentclass[manuscript]{aastex}

\usepackage{txfonts}
\usepackage{longtable}
\usepackage{rotating}
\usepackage{natbib}
\usepackage{graphicx}
\usepackage{graphics}
\usepackage{psfrag}
\usepackage{amssymb}
\bibpunct{(}{)}{;}{a}{}{,}
\def\Teff{$T_{\mathrm{eff}}$}

\def\logR{\ensuremath{\log R^{\prime}_{\rm HK}}}
\def\logRo{\ensuremath{\log R^{\prime (0)}_{\rm HK}}}
\def\logRb{\ensuremath{\log R^{\prime}_{\rm HK (b)}}}
\def\ebv{\ensuremath{E(B-V)}}
\def\bX{{\bf X}}
\def\bx{{\bf x}}
\def\mR{\mathbb{R}}
\def\gpm{$g_{\rm p}^{-1}$}

\shorttitle{A bimodal correlation between stellar chromospheric emission and planet gravity}
\shortauthors{Fossati, Ingrassia \& Lanza}

\begin{document}


\title{A bimodal correlation between host star chromospheric emission and the surface gravity of hot Jupiters}

\author{L.~Fossati\altaffilmark{2}}
\affil{Space Research Institute, Austrian Academy of Sciences, Schmiedlstrasse 		6, A-8042 Graz, Austria}
\email{luca.fossati@oeaw.ac.at}
\and
\author{S.~Ingrassia}
\affil{Department of Economics and Business - University of Catania, Corso Italia, 55, I-95100 Catania, Italy}
\email{s.ingrassia@unict.it}
\and
\author{A.~F.~Lanza}
\affil{INAF-Osservatorio Astrofisico di Catania, Via S.~Sofia, 78, I-95123 Catania, Italy}
\email{nuccio.lanza@oact.inaf.it}

\altaffiltext{2}{Argelander-Institut f\"ur Astronomie der Universit\"at Bonn, 			Auf dem H\"ugel 71, 53121, Bonn, Germany}


%
\begin{abstract}
The chromospheric activity index \logR\ of stars hosting transiting hot Jupiters appears to be correlated with the planets' surface gravity. One of the possible explanations is based on the presence of condensations of planetary evaporated material located in a circumstellar cloud that absorbs the \ion{Ca}{2}\,H\&K and \ion{Mg}{2}\,h\&k resonance line emission flux, used to measure chromospheric activity. A larger column density in the condensations, or equivalently a stronger absorption in the chromospheric lines, is obtained when the evaporation rate of the planet is larger, which occurs for a lower gravity of the planet. We analyze here a sample of stars hosting transiting hot Jupiters tuned in order to minimize systematic effects (e.g., interstellar medium absorption). Using a mixture model, we find that the data are best fit by a two-linear-regression model. We interpret this result in terms of the Vaughan--Preston gap. We use a Monte Carlo approach to best take into account the uncertainties, finding that the two intercepts fit the observed peaks of the distribution of \logR\ for main-sequence solar-like stars. We also find that the intercepts are correlated with the slopes, as predicted by the model based on the condensations of planetary evaporated material. Our findings bring further support to this model, although we cannot firmly exclude different explanations. A precise determination of the slopes of the two linear components would allow one to estimate the average effective stellar flux powering planetary evaporation, which can then be used for theoretical population and evolution studies of close-in planets.
\end{abstract}
%

\keywords{planet-star interactions --- stars: activity --- stars: late-type}

%
\section{Introduction}\label{sec:intro}
There have been many attempts to detect stellar activity excess/deficiency in stars hosting close-in planets (i.e., star--planet interaction; SPI), but with ambiguous results. Works by, e.g., \citet{Shkolniketal03,Shkolniketal05,Shkolniketal08}, \citet{Kashyapetal08}, \citet{Scharf10}, \citet{Gurdemiretal12}, and \citet{Pillitterietal14} reported the detection of a significant increase of stellar activity in stars hosting hot Jupiters (HJs). These findings have been challenged by, e.g., \citet{Poppenhaegeretal10}, \citet{PoppenhaegerSchmitt11}, and \citet{Milleretal15}, who attribute those detections to biases or selection effects.

Based on measurements by \citet{knutson2010}, \citet{Hartman10} discovered a significant correlation between the chromospheric activity index \logR\ of stars hosting transiting HJs and the surface gravity of their planets $g_{\rm p}$. On the other hand, he did not find any correlation of \logR\ with the orbital semi-major axis $a$. This is the only SPI correlation that has been strengthened so far by further dedicated studies \citep{Figueiraetal14}. 

To explain this correlation, \citet{Lanza14} proposed a theoretical model that assumes that the planetary material, evaporated under the action of the stellar extreme ultraviolet (EUV) radiation \citep[][]{Lammeretal03,Lecavelierdesetangsetal04,Lecavelierdesetangs07,sanz-forcada2010,Sanz-Forcadaetal11}, diffuses toward the star along the magnetic field lines of the stellar corona in which the planet is embedded, to finally condense forming prominence-like structures absorbing at the core of the chromospheric resonance lines (e.g., \ion{Mg}{2}\,h\&k, \ion{Ca}{2}\,H\&K). A stronger absorption in the core of the chromospheric resonance lines, where the \logR\ index is measured, occurs when the gravity of the planet is lower, thus making the observed chromospheric emission correspondingly lower.

An independent confirmation of this model comes from the detection of an anomalous lack of activity in the core of the \ion{Ca}{2}\,H\&K and \ion{Mg}{2}\,h\&k resonance lines of WASP-12, hosting an extremely irradiated evaporating hot Jupiter \citep{Fossatietal10,Haswelletal12}. This anomaly is likely caused by circumstellar absorption from material presumably lost by the planet \citep{Fossatietal13}. Numerical simulations of the circumstellar environment in HJs' systems support such a conclusion \citep[e.g.,][]{Cohenetal11,Matzakosetal15}. In addition, observations have shown that atmospheric evaporation can be conspicuous and it is a key factor shaping planet structure, evolution, and circumplanetary environment for both hot Jupiters \citep[e.g.,][]{vidal-madjar2003} and close-in low-mass planets \citep{kulow2014,ehrenreich2015}.

A limitation of the model proposed by \citet{Lanza14} is that the best fit obtained when it is applied to the observations has a reduced $\chi^2$ remarkably greater than unity. In the present Letter, we significantly improve the best fit by treating the observed sample of stars as a mixture of two different subsets with intrinsic chromospheric activity above or below the so-called Vaughan--Preston gap \citep[hereafter VP;][]{VaughanPreston80} observed for solar-like stars \citep[see][]{Wright04,Grayetal06}. 
\section{Observations}\label{sec:observations}
We consider the sample of late-type stars with transiting planets investigated by \citet{Figueiraetal14} and select the objects with effective temperature 4200$\leq$\Teff$\leq$6200\,K, where the \logR\ index is best calibrated. We further restrict the sample to the systems with semi-major axis $a\leq$0.1\,AU and $M_{\rm p}$$\geq$0.1\,$M_{\rm J}$ ($M_{\rm p}$ is the planetary mass and $M_{\rm J}$ Jupiter's mass) and consider only the brightest stars ($V$\,$\leq$\,13\,mag) with a color excess \ebv$\leq$0.06\,mag, as derived from extinction maps by \citet{AmoresLepine05}, in order to reduce systematic effects caused by interstellar absorption. We excluded known multiple systems because the presence of several (close-in) planets may further complicate the correlations we want to investigate. 

Jupiter-like planets are expected to have a hydrogen-dominated atmosphere, thus their evaporation is powered by the stellar EUV radiation ($\lambda<$912\,\AA). Smaller planets can have instead a different atmospheric composition, possibly implying a different effective passband for the UV flux powering evaporation. It is to exclude those planets that we limit our sample to $M_{\rm p}\geq$\,0.1\,$M_{\rm J}$. 

Typical standard deviations in \logR\ are about 0.09\,dex \citep{Lanza14}. The standard deviation of the planetary gravity is computed propagating the uncertainties in the system parameters (from exoplanets.org). The considered sample of stars is listed in Table~\ref{tab:stars}.
\begin{table}[ht]
\vspace{-3.6cm}
\caption[ ]{Planetary systems considered in this work. Column two lists the measured \logR\ values from \citet{Figueiraetal14}. The following columns give the planets' surface gravity (in cgs), the relative uncertainty, derived from the planets' mass and radius, the semi-major axis, the stellar mass, and the planetary mass (data from exoplanet.org). The systems marked with a star (*) have been removed from the sample (see Sect.~\ref{sec:results}).}
\label{tab:stars}
\begin{center}
\begin{scriptsize}
\begin{tabular}{l|ccc|ccc}
\hline
\hline
System & \logR & log\,$g_{\rm p}$ & $\sigma$log\,$g_{\rm p}$ & $a$  & $M_{\rm s}$ & $M_{\rm p}$ \\
       &       &                  &                          &(AU)&($M_{\odot}$)&($M_{\rm J}$)\\
\hline
CoRoT-2		& $-$4.331 & 3.582 & 0.026 & 0.0281  & 0.97$\pm$0.06 & 3.31$\pm$0.16 \\
HAT-P-1		& $-$4.984 & 2.874 & 0.020 & 0.05561 & 1.15$\pm$0.05 & 0.52$\pm$0.02 \\
HAT-P-3		& $-$4.904 & 3.331 & 0.056 & 0.03866 & 0.92$\pm$0.03 & 0.59$\pm$0.02 \\
HAT-P-4		& $-$5.082 & 3.019 & 0.041 & 0.0446  & 1.26$\pm$0.14 & 0.68$\pm$0.04 \\
HAT-P-5		& $-$5.061 & 3.224 & 0.053 & 0.04079 & 1.16$\pm$0.06 & 1.06$\pm$0.11 \\
HAT-P-12	& $-$5.104 & 2.753 & 0.030 & 0.0384  & 0.73$\pm$0.02 & 0.21$\pm$0.01 \\
HAT-P-13	& $-$5.138 & 3.109 & 0.054 & 0.0426  & 1.22$\pm$0.10 & 0.85$\pm$0.04 \\
HAT-P-16	& $-$4.862 & 3.796 & 0.043 & 0.0413  & 1.22$\pm$0.04 & 4.19$\pm$0.09 \\
HAT-P-17	& $-$5.039 & 3.110 & 0.028 & 0.0882  & 0.86$\pm$0.04 & 0.53$\pm$0.02 \\
HAT-P-27	& $-$4.785 & 3.169 & 0.057 & 0.0403  & 0.92$\pm$0.06 & 0.66$\pm$0.03 \\
HAT-P-31	& $-$5.312 & 3.673 & 0.139 & 0.055   & 1.22$\pm$0.06 & 2.17$\pm$0.10 \\
HAT-P-44	& $-$5.247 & 2.773 & 0.070 & 0.0507  & 0.94$\pm$0.04 & 0.39$\pm$0.03 \\
HD\,149026	& $-$5.030 & 3.233 & 0.449 & 0.04288 & 1.30$\pm$0.10 & 0.36$\pm$0.01 \\
HD\,189733	& $-$4.501 & 3.338 & 0.056 & 0.03142 & 0.80$\pm$0.40 & 1.14$\pm$0.02 \\
HD\,209458	& $-$4.970 & 2.968 & 0.015 & 0.04747 & 1.15$\pm$0.02 & 0.71$\pm$0.02 \\
TrES-1		& $-$4.738 & 3.194 & 0.038 & 0.0393  & 0.88$\pm$0.07 & 0.76$\pm$0.05 \\
TrES-2		& $-$4.949 & 3.342 & 0.025 & 0.03556 & 0.98$\pm$0.06 & 1.25$\pm$0.05 \\
TrES-3		& $-$4.549 & 3.444 & 0.058 & 0.0226  & 0.92$\pm$0.04 & 1.91$\pm$0.07 \\
WASP-2		& $-$5.054 & 3.256 & 0.034 & 0.03138 & 0.84$\pm$0.11 & 0.85$\pm$0.04 \\
WASP-4		& $-$4.865 & 3.197 & 0.024 & 0.02312 & 0.93$\pm$0.05 & 1.24$\pm$0.06 \\
WASP-5		& $-$4.720 & 3.472 & 0.045 & 0.02729 & 1.00$\pm$0.06 & 1.64$\pm$0.08 \\
WASP-11		& $-$4.823 & 3.019 & 0.037 & 0.0439  & 0.82$\pm$0.03 & 0.46$\pm$0.03 \\
WASP-13		& $-$5.263 & 2.810 & 0.055 & 0.05379 & 1.09$\pm$0.05 & 0.48$\pm$0.05 \\
WASP-16		& $-$5.100 & 3.320 & 0.063 & 0.0421  & 1.02$\pm$0.10 & 0.86$\pm$0.06 \\
WASP-19		& $-$4.660 & 3.152 & 0.021 & 0.01616 & 0.90$\pm$0.04 & 1.11$\pm$0.04 \\
WASP-22		& $-$4.900 & 3.036 & 0.030 & 0.04698 & 1.10$\pm$0.30 & 0.59$\pm$0.02 \\
WASP-23		& $-$4.680 & 3.368 & 0.061 & 0.0376  & 0.78$\pm$0.13 & 0.87$\pm$0.09 \\
WASP-26		& $-$4.980 & 3.194 & 0.049 & 0.03985 & 1.12$\pm$0.03 & 1.03$\pm$0.02 \\
WASP-41		& $-$4.670 & 3.193 & 0.056 & 0.04    & 0.95$\pm$0.09 & 0.92$\pm$0.07 \\
WASP-42		& $-$4.900 & 3.026 & 0.052 & 0.0458  & 0.88$\pm$0.08 & 0.50$\pm$0.03 \\
WASP-48		& $-$5.135 & 2.946 & 0.054 & 0.03444 & 1.19$\pm$0.05 & 0.99$\pm$0.09 \\
WASP-50		& $-$4.670 & 3.441 & 0.028 & 0.02913 & 0.86$\pm$0.06 & 1.44$\pm$0.07 \\
WASP-52*	& $-$4.400 & 2.847 & 0.027 & 0.0272  & 0.87$\pm$0.03 & 0.46$\pm$0.02 \\
WASP-58*	& $-$4.400 & 3.071 & 0.115 & 0.0561  & 0.94$\pm$0.10 & 0.89$\pm$0.07 \\
WASP-59*	& $-$4.100 & 3.550 & 0.073 & 0.06969 & 0.72$\pm$0.03 & 0.86$\pm$0.04 \\
WASP-69*	& $-$4.540 & 2.823 & 0.044 & 0.04525 & 0.83$\pm$0.03 & 0.30$\pm$0.02 \\
WASP-70		& $-$5.230 & 3.366 & 0.098 & 0.04853 & 1.11$\pm$0.04 & 0.60$\pm$0.02 \\
WASP-84		& $-$4.430 & 3.287 & 0.026 & 0.0771  & 0.84$\pm$0.04 & 0.69$\pm$0.03 \\
WASP-117	& $-$4.950 & 2.815 & 0.057 & 0.09459 & 1.13$\pm$0.29 & 0.27$\pm$0.01 \\
XO-1		& $-$4.958 & 3.202 & 0.043 & 0.0488  & 1.00$\pm$0.03 & 0.90$\pm$0.07 \\
XO-2		& $-$4.988 & 3.210 & 0.029 & 0.0369  & 0.98$\pm$0.02 & 0.62$\pm$0.02 \\
\hline
\end{tabular}
\end{scriptsize}
\end{center}
\end{table}
%
\section{Model}\label{sec:model}
\subsection{Expected Correlations between Stellar Chromospheric Emission and Planet Gravity}\label{sec:model_correlations}
The model proposed by \citet{Lanza14} assumes that the orbital motion of the planet induces a steady energy dissipation driving the host star's coronal field toward a minimum energy state characterized by mostly closed field lines that extend up to a few tens of stellar radii \citep{Lanza09}, within HJs' orbital distance. Among the possible configurations, there are some with a dip along the field lines, close to the star, where the matter evaporating from the planet can radiatively cool, condense, and be stably supported against the stellar gravity. Those condensations are expected to be similar to solar prominences with a typical temperature of $\sim$10$^4$\,K and an electron density of $n_{\rm e}\sim$10$^{10}$\,cm$^{-3}$, and absorb in the chromospheric resonance lines, notably \ion{Ca}{2}\,H\&K and \ion{Mg}{2}\,h\&k. 

Assuming that the evaporation is powered by the stellar EUV flux and that the evaporated matter diffuses along the field lines reaching the condensation site close to the star, the model predicts that \citep{Lanza14}
\begin{equation}
\log R^{\prime}_{\rm HK}=\log R^{\prime (0)}_{\rm HK} - \gamma g_{\rm p}^{-1}\,,
\label{regression}
\end{equation}
where \logR\ is the measured chromospheric index of the star, \logRo\ is its intrinsic index (i.e., without any absorption by the circumstellar material), and the slope $\gamma$ is given by
\begin{equation}
\gamma=0.0434\,\frac{\alpha \, \eta \, F_{\rm EUV}}{m_{\rm p} \, c_{\rm s}},
\label{eq2}
\end{equation}
where $\alpha$ is the line absorption coefficient per hydrogen atom, $F_{\rm EUV}$ the EUV stellar flux powering evaporation, $\eta$ the average heating efficiency, $m_{\rm p}$ the proton mass, and $c_{\rm s}$ the sound speed of the evaporating plasma. 

Stars hosting transiting HJs are generally not biased against solar-amplitude stellar activity. We therefore expect that the \logRo\ distribution follows the distribution found for solar-like stars. This can be approximated as the sum of two Gaussians with mean values of about $-$4.9 and $-$4.5 and equal standard deviations of $\approx0.25$\,dex \citep{Wright04,Grayetal06}. The former corresponds to low-activity stars and the latter to high-activity stars, i.e. below and above the VP gap, respectively. The VP gap seems not to be present at low metallicity \citep{Grayetal06}, but our stars have about solar metallicity. We expect to find an imprint of such a bimodal distribution in Eq.~\ref{regression}: \logRo\ is equal to the unabsorbed value of the chromospheric index and $\gamma$ is proportional to the flux emitted in the EUV passband, which is greater for higher-activity stars, hence \logRo\ and $\gamma$ are correlated.

\citet{Sanz-Forcadaetal11} suggested that $F_{\rm EUV} \propto F_{\rm X}^{0.860\pm0.073}$, where $F_{\rm X}$ is the X-ray coronal flux in the passband $5-100$\,\AA, roughly coincident with that of PSPC on board of ROSAT used by \citet{Pitersetal97} to correlate $F_{\rm X}$ with the excess chromospheric emission $\Delta F_{\rm HK}$ of late-type stars. The excess emission $\Delta F_{\rm HK}$ is obtained from the measured chromospheric flux by subtracting the so-called basal flux \logRb, the minimum emission flux present for stars with a chromosphere, which for main-sequence stars should be\footnote{The \logRb\ value has a slight dependence on the effective temperature/mass.} $-$5.1 \citep{Wright04}. According to \citet{Pitersetal97}, $F_{\rm X} \propto (\Delta F_{\rm HK})^{2.1\pm0.2}$; therefore, we expect $F_{\rm EUV} \propto (\Delta F_{\rm HK})^{1.82\pm0.33}$. By introducing the chromospheric index, this can be written as
\begin{equation}
F_{\rm EUV} \propto \left[ 10^{\log R^{\prime (0)}_{\rm HK}} - 10^{\log R^{\prime}_{\rm HK (b)}} \right]^{\,\beta}\,,
\label{eq3}
\end{equation}
where $\beta=1.82\pm0.33$. The slope $\gamma$ of \logR\ vs. \gpm, as given by Eq.~\ref{eq2}, is therefore expected to be correlated with the intercept \logRo\ of the same line. The heating efficiency $\eta$ and the sound speed $c_{\rm s}$ in Eq.~\ref{eq2} are almost constant in the EUV flux regime characteristic of our stars \citep{Murray-Clayetal09,Shematovichetal14}; thus, we predict that the ratio $\gamma / F_{\rm EUV}$, where $F_{\rm EUV}$ can be estimated from Eq.~\ref{eq3}, is almost constant.

The slopes of the regression lines (Eq.~\ref{regression}) computed for stars above and below the VP gap are therefore expected to be correlated to their intercepts \logRo, while the intercepts themselves are expected to be approximately $-4.5$ and $-4.9$, respectively, where we have assumed that all the other parameters entering in $\gamma$ are the same for all stars.

The model assumes that the condensations have a cylindrical symmetry around the axis perpendicular to the planet's orbital plane; hence, it does not predict any dependence of the absorption on the planet's orbital phase, despite that single condensations may be discrete objects. For WASP-12, the line core absorption is indeed always present, regardless of the planet's orbital phase.
\subsection{Applying a Cluster-weighted Model (CMW) to Fit the Data}\label{sec:fit2data}
Our stars cannot be immediately assigned to the groups above and below the VP gap because of the circumstellar/ISM absorption, intrinsic variability, and statistical measurement errors. Therefore, in order to proceed in the data analysis, we fit the \logR--\gpm\ correlation by considering a statistical mixture model consisting of a combination of different linear regressions called CWM or {\em Mixtures of regressions with random covariates} \citep[see][]{Ingrassiaetal2012}. Assume we are provided with a random pair $(\bX',\,Y)'$ having joint density $p(\bx,\,y)$, where $Y$ is a real-valued response variable and $\bX $ is a vector of covariates with values in $\mR^p$. In the CWM framework, the joint density of each mixture component can be factorized into the product of the conditional density of $Y|\bX=\bx$ and the marginal density of $\bX$ by assuming a parametric functional dependence of $Y$ on $\bx$.

The overall formulation is
\begin{equation}
p(\bx,\,y;\,\psi)\,=\,\sum_{j=1}^{J}\pi_j\,p(y|\bx,j)\,p(\bx|j)\,, 
\label{eq:CWM base} 
\end{equation}
where $J$ indicates the number of components in the model, $p(y|\bx,j)$ is the component conditional density of the response variable $Y$ given, $\bx$, $p(\bx|j)$ is the component marginal density of $\bX$, and $\pi_j$ is the mixture weight for component $j$ (defined so that $\pi_j>0$ and $\sum_j \pi_j=1$). As in \citet{Ingrassiaetal2012,Ingrassiaetal2014}, we assume the component densities, $p(\bx|j)$ and $p(y|\bx,j)$, to be Gaussian or Student-$t$ distributed. We note that the Student-$t$ distribution provides more robust fitting for groups of observations with longer than normal tails or noisy data.

The independent variable $X$ is \gpm, while the response variable is \logRo. Assuming Gaussian distributions, Eq.~\ref{eq:CWM base} becomes
\begin{equation}
p(g_{\rm p}^{-1},\log R^{\prime (0)}_{{\rm HK}}; \psi)=\sum_{j=1}^{J}\pi_j\,\phi(\log R^{\prime (0)}_{{\rm HK}}, \log R^{\prime (0)}_{{\rm HK} j}-\gamma_{j} g_{\rm p}^{-1}, \sigma^{2}_{j})\,\,\phi(g_{\rm p}^{-1}; \mu_{g,j},\sigma^2_{g,j}), 
\label{eq:CWMRg} 
\end{equation}
where $\log R^{\prime (0)}_{{\rm HK} j}$ is the intercept and $\gamma_{j}$ the slope of the $j$th regression considered in the mixture, while $\mu_{g,j},
\sigma^2_{g,j}$ are the parameters of the distribution of \gpm. Finally, the vector ${\bf \psi}$ is the vector of all parameters of the mixture model. In other words, we assume that the observed \logR--\gpm\ correlation comes from a superposition of different linear regressions weighted on both local densities $\phi(g^{-1}; \mu_{g,j}, \sigma^2_{g,j})$ and mixing weights $\pi_{j}$. 

For a given number of components $J$, the determination of the best-fit parameters ${\bf \psi}$ and of the probability of assignment of each data point $(\log R^{\prime}_{\rm HK}, g_{\rm p}^{-1})$ to a component of the mixture is performed by means of a maximum-likelihood approach based on the Expectation--Maximisation algorithm \citep[see][]{Dempsteretal1977,Ingrassiaetal2014}. The selection of the best number of components $J$ can be performed by comparing the likelihoods of the models obtained for different $J$s, simultaneously allowing for a penalty factor that discourages the selection of models with a greater number of parameters. We take into account the Bayesian information criterion (BIC), usually adopted in the mixture models literature \citep{McLaPe:00}, to guide the model selection. Once the parameters have been estimated, the data points are classified in either group according to the maximum a posteriori probability criterion.
\section{Results}\label{sec:results}
\begin{figure}
\centerline{\includegraphics[width=14.5cm,clip]{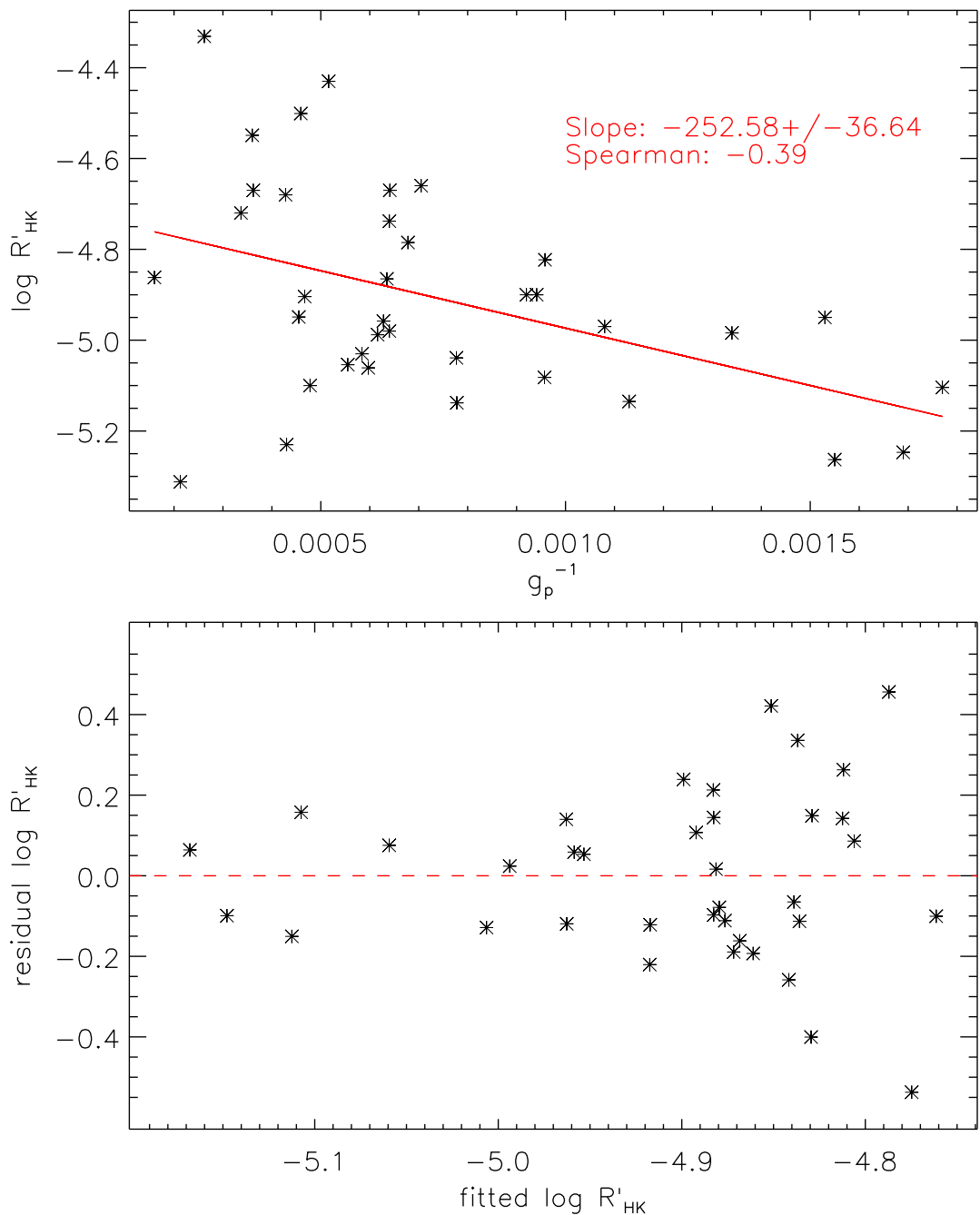}}
\caption{Top: one-component fit to the \logR--\gpm\ correlation. Slope and Spearman-rank correlation coefficient are given on the top right corner. Bottom: residuals from a one-component linear fit to the data. The distribution of the residuals depends upon the \logR\ value, hence a one-component model is not a good fit.}
\label{fig:residuals}
\end{figure}

\begin{figure}
\centerline{\includegraphics[width=\hsize,clip]{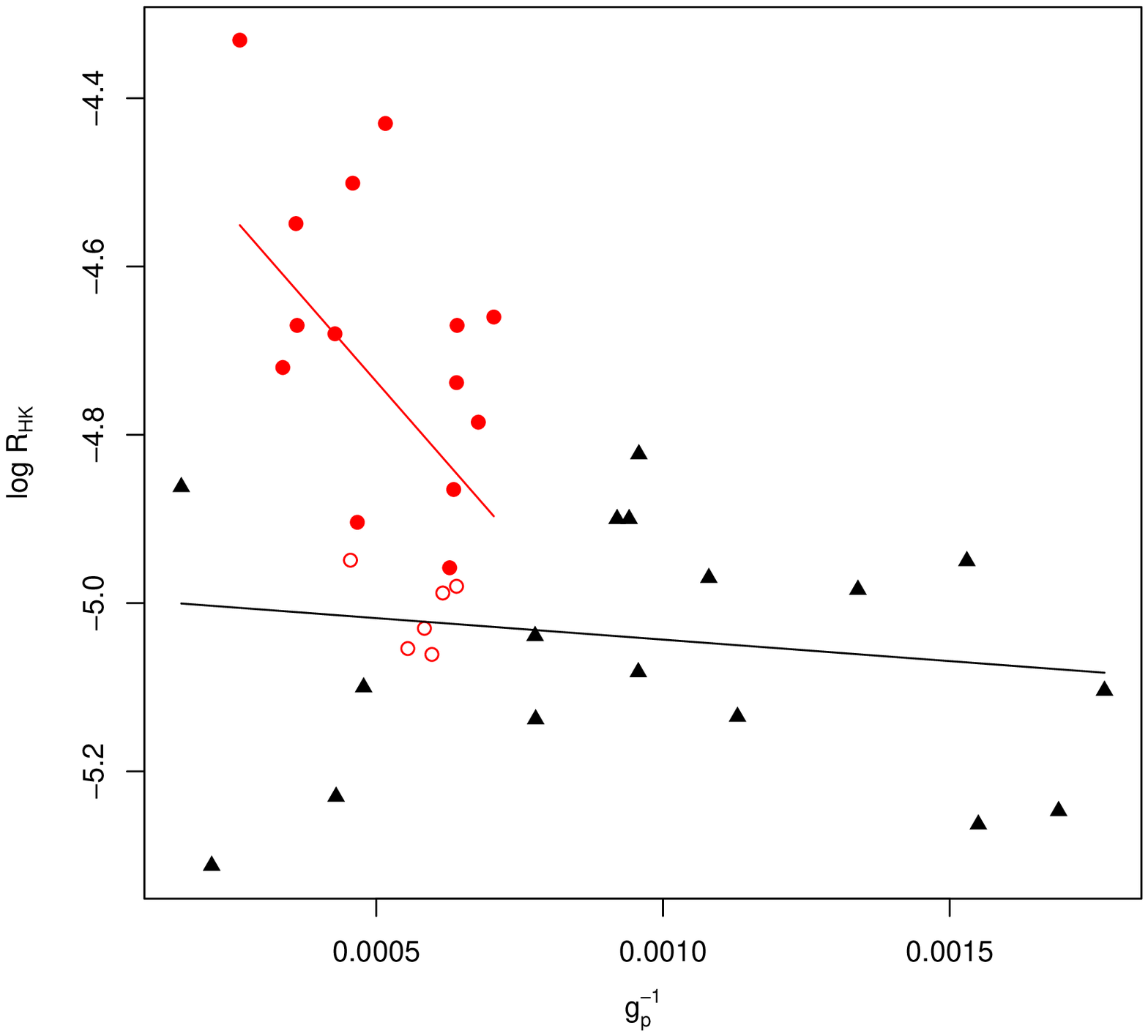}}
\caption{Chromospheric emission index \logR\ vs. the inverse of the planet gravity \gpm\ (in cm$^{-1}$\,s$^{2}$) with the two best-fit regression lines of our mixture model in black and red. The data points assigned to each of the two regressions are plotted with the same color coding of the corresponding regression line. The open circles indicate points belonging to the high-activity component with an a posteriori probability between 0.5 and 0.65, hence points which may also belong to the low-activity component.}
\label{fig:2regressions}
\end{figure}

For our sample, we selected a mixture model with $J=2$ components. Based on the BIC values, the models with one and two components are equivalent, but the large heterogeneity of the data (the residuals depend upon the \logR\ value; Fig.~\ref{fig:residuals}) and the large $\chi^2$ \citep{Lanza14} make the one-component solution less likely. Following the BIC criterion, the two-component solution is more likely than the three-component one. The corresponding two-component linear regression lines are plotted in Fig.~\ref{fig:2regressions}. Four data points\,/\,systems (WASP-52, WASP-58, WASP-59, and WASP-69) have been discarded because lying far above both regression lines, the measured \logR\ could be affected by flaring or the stars could be caught during a phase of enhanced activity \citep[e.g., owing to SPI;][]{Shkolniketal08,Lanza13}. 

We find 17 points in the low-activity component and 20 in the high-activity component with intercepts $\log R^{\prime (0)}_{\rm HK1}=-4.84\pm0.04$ and $\log R^{\prime (0)}_{\rm HK2}=-4.41\pm0.15$, and slopes $\gamma_{1}=-153\pm36$ and $\gamma_{2}=-706\pm270$\,cm\,s$^{-2}$, respectively. There are systems/points that could belong to both components, but the results are not affected by considering these points belonging to either of them. As for the correlation analysis between \logR\ and \gpm\ in either group, in this framework, usual goodness-of-fit statistics for bivariate data such as $R^2$ cannot be used. New and suitable diagnostic statistics are currently under development.

\begin{figure}
\centerline{\includegraphics[width=13.5cm,clip]{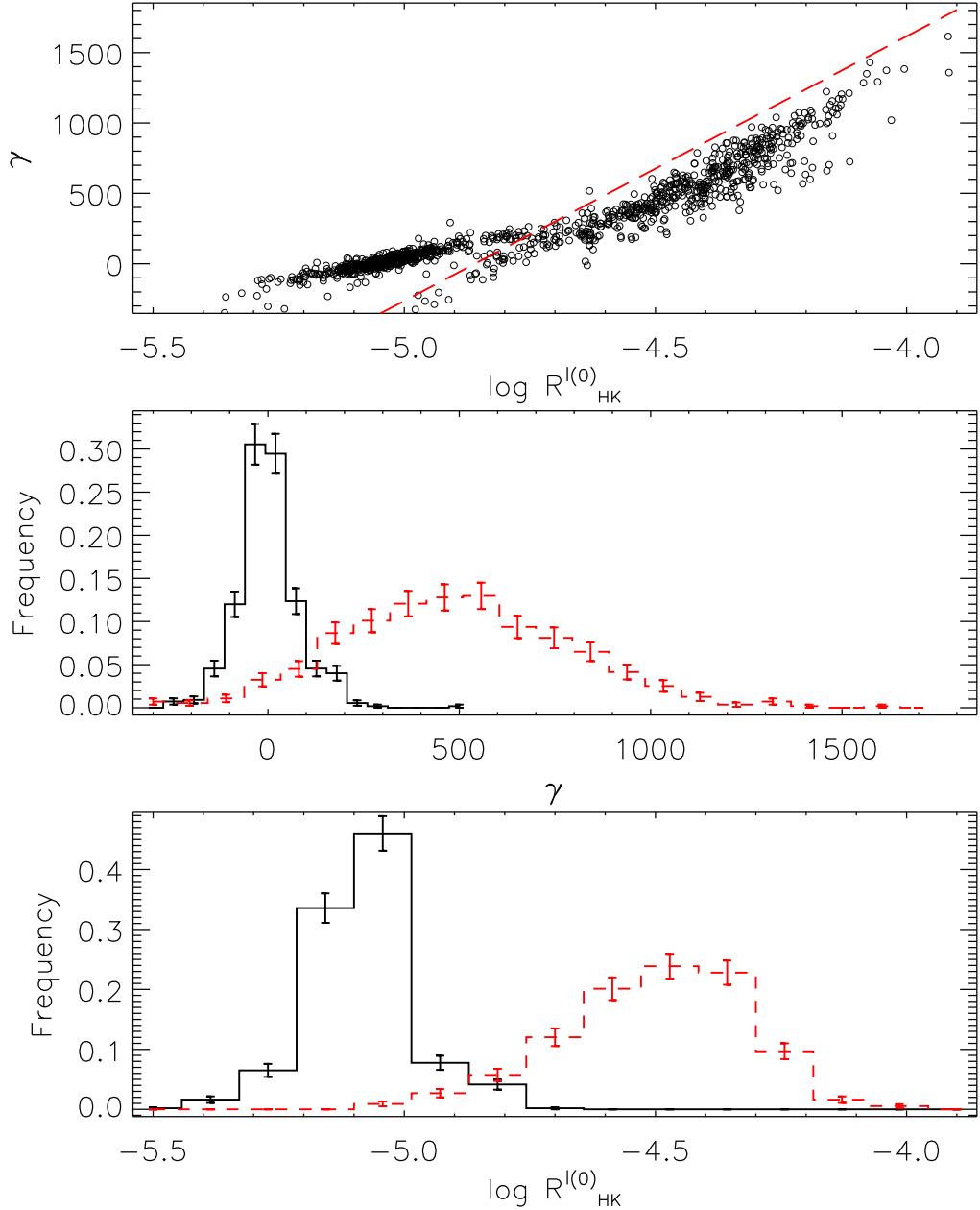}}
\caption{Top: slope $\gamma$ vs. intercept \logRo\ for the regression lines found by fitting our two-component mixture model to the 1000 mock data sets obtained to study the impact of the statistical errors on the parameters of our model. The red dashed line marks the boundary separating the joint distribution of the two regression lines, thus splitting the whole sample into two distinct clusters. Middle: distributions of the slopes $\gamma$ of the points in the top panel belonging to the clusters above (black solid line) and below (red dashed line) the line marking the boundary between the two clusters. Bottom: same as the middle panel, but for the intercepts \logRo.}
\label{fig:correlation}
\end{figure}

The impact of the statistical errors is quantified as follows. We generate 10$^4$ mock data sets by adding to the coordinates of each point of the true data set random normally distributed deviates with standard deviations $\sigma_{\log R^{\prime}_{\rm HK}}=0.09$ and $\sigma_{g_{\rm p}^{-1}}$ (Table~\ref{tab:stars}), respectively, to simulate the effect of the errors. We apply our mixture model to the mock data sets; we then consider the 1000 data sets returning the highest likelihood values and, for the regression lines found by fitting at least 11 points\footnote{This value has been chosen in order to obtain about the same number of (\logRo,$\gamma$) combinations belonging to each of the two components, not to affect the statistics. A different choice, e.g. 9--13 points, would affect neither the results nor the position of the boundary line shown in Fig.~\ref{fig:correlation}.}, we analyze the a posteriori joint distribution of the intercept \logRo\ and the slope $\gamma$ (Fig.~\ref{fig:correlation}).

Figure~\ref{fig:correlation} shows two clusters of points, separated by the red dashed line that we associate to each component of our mixture model. This indicates that the statistical errors on the two variables \logR\ and \gpm\ are small enough to allow us to clearly identify the two components of the mixture. The dispersion of the points around the barycenter of each cluster provides a measure of the impact of the statistical errors on the parameters. The fraction of points in Fig.~\ref{fig:correlation} belonging to the two clusters is 49.8\% and 50.2\%, indicating that each component has been sampled virtually with the same a posteriori probability. 

We consider the distributions of the intercepts and of the slopes separately for the points belonging to the two clusters and use them to estimate the statistical uncertainties on the intercepts and slopes of the two regression lines of our mixture model (Fig.~\ref{fig:correlation}). The average values of the intercepts and their standard deviations for the points belonging to the two clusters are \logRo=$-$5.03$\pm$0.10 and \logRo=$-$4.44$\pm$0.18, which are closer than $\log R^{\prime (0)}_{\rm HK 1,2}$ to the values expected as a consequence of the bimodal distribution of the intrinsic chromospheric emission in late-type stars \citep{Wright04,Grayetal06}. The mean intercept of the lower-activity cluster is still compatible with $\log R^{\prime (0)}_{\rm HK1}$, although its derived error was $\approx$2.5 times smaller than that estimated a posteriori with our method. For the higher-activity component, the agreement with $\log R^{\prime (0)}_{\rm HK2}$ has improved, both in terms of estimated mean value and uncertainty. This shows that thoroughly taking into account the statistical uncertainties that affect the data set strengthens the prediction of a bimodal distribution of the intrinsic chromospheric emission.

The average values of the slopes $\gamma$ and their standard deviations are 23.0$\pm$87.0 and 537.2$\pm$310.3\,cm\,s$^{-2}$ for the two clusters in Fig.~\ref{fig:correlation}. The cluster with the greater average intercept has the steeper mean slope, as expected on the basis of our model. However, the average slope of the low-activity cluster is severely affected by the statistical uncertainties and its value is not significantly different from zero, showing that the slope of the regression of the low-activity component is the most uncertain parameter in our mixture model. It may well be that the correlation disappears once the host star has reached an activity below a certain level in which case the EUV flux may not be high enough to produce condensations that appreciably absorb the chromospheric flux or the stellar magnetic field, generating the coronal loops, not be strong enough.

We calculated then the $R$=$\gamma/F_{\rm EUV}$ ratios to check whether the values obtained from the high- and low-activity components are compatible, as expected on the basis of our model. We estimated the stellar EUV flux using $\beta$=1.82 in Eq.~\ref{eq3} and an activity basal level of \logRb=$-$5.1 \citep{Wright04}. We obtained $R_1$=5.0$\pm$2.7$\times$10$^{11}$ and $R_2$=2.0$\pm$1.6$\times$10$^{11}$, in agreement within 1$\sigma$ with our prediction of a constant ratio $\gamma/F_{\rm EUV}$.
\section{Discussion}\label{conclusions}
%
\begin{figure}
\vspace{-2.0cm}
\centerline{\includegraphics[angle=-90,width=12.cm,clip]{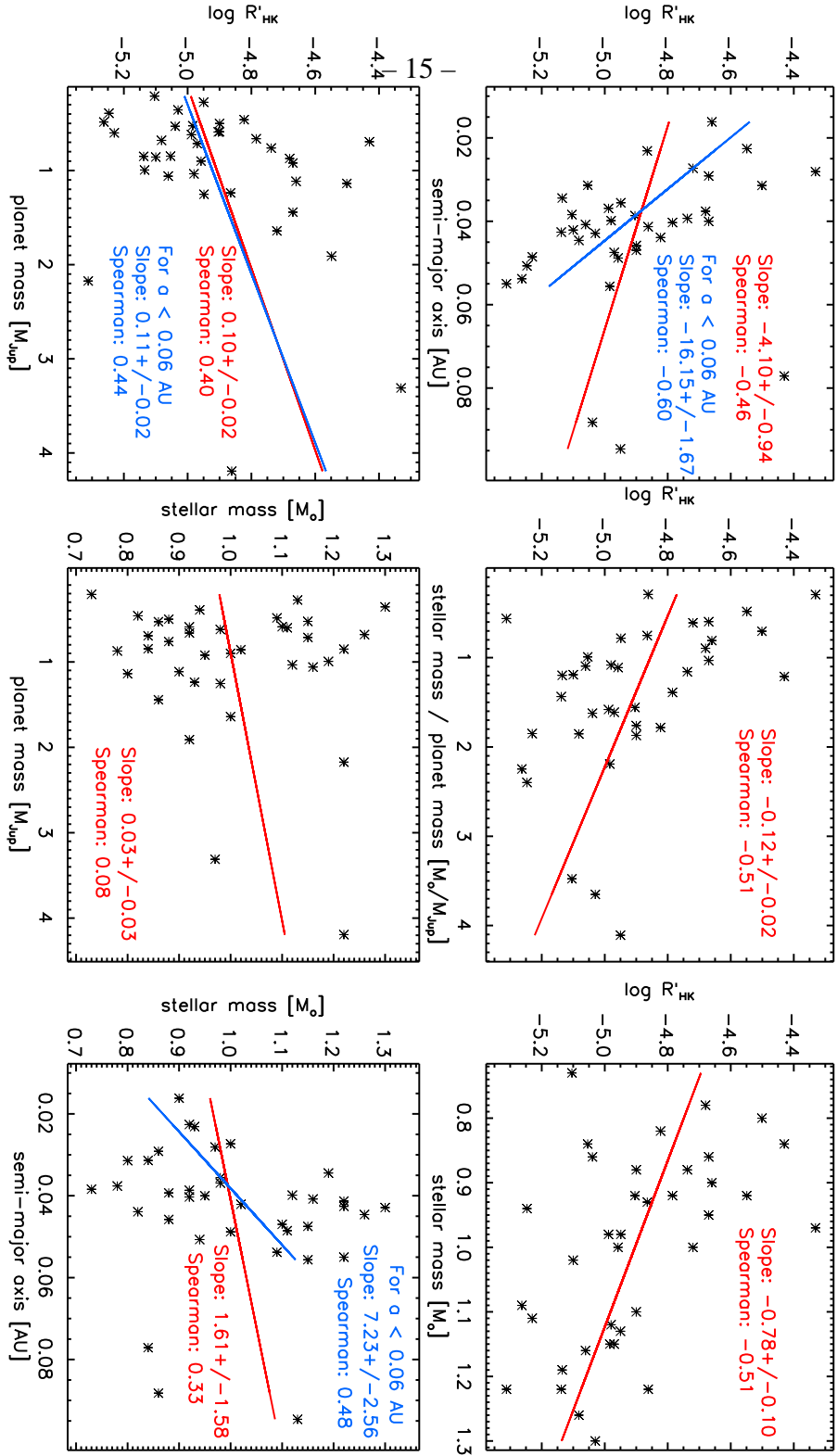}}
\caption{Top: correlations of the \logR\ values with the orbit semi-major axis (left), the stellar-to-planetary mass ratio (middle), and the stellar mass $M_{\rm s}$ (right). Bottom: correlation of the \logR\ values with planetary mass $M_{\rm p}$ (left) and correlations of $M_{\rm s}$ with $M_{\rm p}$ (middle) and semimajor axis (right). We also considered correlations for $a<$0.06\,AU. Each panel gives, for each correlation, the value of the slope and of the Spearman-rank correlation coefficient.}
\label{fig:multiplot}
\end{figure}

\citet{Hartman10} already discussed the possibility that the \logR--\gpm\ correlation is caused by SPI. Because of our refined data set, we reanalyse here the possible role of SPI.

The top panels of Fig.~\ref{fig:multiplot} show that \logR\ is correlated with the semimajor axis \citep[in contrast to what was found by][]{Hartman10}, the stellar-to-planetary mass ratio, and the stellar mass: a lower activity is found for systems with a higher stellar-to-planetary mass ratio hosting planets with wider orbits. The \logR--$a$ correlation becomes even more significant when taking into account systems with $a<$0.06\,AU, directly pointing toward an SPI origin of the \logR--\gpm\ correlation.

The bottom panels of Fig.~\ref{fig:multiplot} clarify instead that these correlations may not be caused by SPI. We find no relation between planetary and stellar mass and a weak \logR--$M_{\rm p}$ correlation; the latter does not change when considering only the most close-in planets and weakens slightly (Spearman-rank coefficient of 0.38) when considering planets with M$_{\rm p}>$0.3\,$M_{\rm J}$ in order to avoid the biases described by \citet{PoppenhaegerSchmitt11}. This shows that the \logR--$M_{\rm s}$/$M_{\rm p}$ correlation may be a consequence of the strong \logR--$M_{\rm s}$ correlation. Supporting this interpretation, \citet{canto2011}, using systems hosting planets discovered by radial velocity, found a weak \logR--$M_{\rm p}$ correlation with a 30\% significance.

We find a negative \logR--$M_{\rm s}$ correlation. If one considers $B-V$ or effective temperature as a proxy for stellar mass, this contrasts with what typically found for large samples of middle-aged field stars \citep[e.g.,][]{Grayetal06,Fossatietal13,pace2013,Milleretal15}. In addition, \citet{canto2011}, using a large sample of non-transiting systems, found a positive \logR--$M_{\rm s}$ correlation. We argue that this difference is due to the fact that our sample of stars hosts transiting planets: condensations of planetary evaporated material would lie mostly close to the orbital plane and, on average, higher-mass stars have a higher EUV flux, which leads to a higher planet evaporation rate.

We find also a significant correlation between $M_{\rm s}$ and $a$, which is probably connected to ground-based detection biases of transiting planets: the ground-based detection probability of a planetary transit depends upon the orbital period, which is directly proportional to $\sqrt{a^3/{\rm M_s}}$. Since the orbital period of hot Jupiters falls in a restricted range (1--5\,days) peaking at $\approx3$ days, the $M_{\rm s}$--$a$ correlation reflects the fact that the ratio $a^3$/$M_{\rm s}$ has to be about constant.

The \logR--$a$ correlation may therefore find its origin in the $M_{\rm s}$--$a$ and \logR--$M_{\rm s}$ correlations. We note that these parameters are affected by various uncertainties, which may for example explain the difference in the significance between the \logR--$a$ and $M_{\rm s}$--$a$ correlations.

We find no reason to conclude that the \logR--\gpm\ correlation is caused by SPI, although we are not able to exclude it completely. This conclusion is supported by results obtained from similar studies of X-ray fluxes and \logR\ values of various samples of planet hosting stars \citep{Poppenhaegeretal10,PoppenhaegerSchmitt11,Milleretal15}. The scenario proposed by \citet{Lanza14} remains therefore the most likely explanation for the \logR--\gpm\ correlation and our results support it. It is left to future observations (particularly aiming at increasing the sample of stars and further removing the systematic biases, e.g. interstellar absorption) and modeling to support or contradict this interpretation.
\section{Conclusions}
We found evidence of a mixture of two distributions in the correlation between the chromospheric index \logR\ and the surface gravity $g_{\rm p}$ of HJs \citep{Hartman10,Figueiraetal14}. Specifically, the BIC criterion and considerations on the residuals from a single linear model give a strong preference to a two-component mixture model.

\citet{Lanza14} proposed that the \logR--$g_{\rm p}$ correlation may be due to planetary evaporated material that absorbs in the chromospheric resonance lines, where \logR\ is measured, in connection to the fact that planets with a lower surface gravity have a higher evaporation rate. Within this context, the two-component mixture model is naturally interpreted as a consequence of the VP gap in the distribution of the \logR\ values in late-type stars. In addition, we found that the intercepts are correlated with the slopes, as predicted by the model. 

We re-evaluate the role of SPI in the \logR--$g_{\rm p}$ correlation finding evidence favoring more the model based on the condensations of planetary evaporated material rather than SPI, though the evidence is not strong enough to completely exclude it. Following the model proposed by \citet{Lanza14}, a precise determination of the slopes $\gamma$ of the two linear components would allow one to estimate the average $\eta \times F_{\rm EUV}$ (assuming the sound speed does not vary dramatically from one system to the other), the effective stellar flux powering planetary evaporation, which can then be used for theoretical population and evolution studies of close-in planets.
\clearpage
\acknowledgments
L.F. acknowledges financial support from the Alexander von Humboldt foundation. A.F.L. acknowledges support by INAF through the Progetti Premiali funding scheme of the Italian Ministry of Education, University, and Research. We thank the anonymous referee for the useful comments. This research has made use of the Exoplanet Orbit Database and the Exoplanet Data Explorer at exoplanets.org.

\end{document}